\documentclass[twocolumn,secnumarabic,amssymb, nobibnotes, aps, prl]{revtex4-2}

\usepackage[dvipdfmx]{graphicx}
\usepackage{color}
\usepackage{amsmath}

\begin{document}

\title{Experimental demonstration of attosecond hard X-ray pulses}

\author{Ichiro Inoue$^{1,2,3}$} 
\email{ichiro.inoue@edu.k.u-tokyo.ac.jp}
\author{River Robles$^{4,5,6,*}$} 
\author{Aliaksei Halavanau$^4$} 
\email{aliaksei@slac.stanford.edu}
\author{Veronica Guo$^{4,5}$}
\author{Thomas M. Linker$^{6,7}$} 
\author{Andrei Benediktovitch$^{8}$} 
\author{Stasis Chuchurka$^{8}$} 
\author{Matthew H. Seaberg$^{4}$}
\author{Yanwen Sun$^{4}$}
\author{Diling Zhu$^{4}$}
\author{David Cesar$^{4}$}
\author{Yuantao Ding$^{4}$}
\author{Vincent Esposito$^{4}$}
\author{Paris Franz$^{4,5}$}
\author{Nicholas S. Sudar$^{4}$}
\author{Zhen Zhang$^{4}$}
\author{Taito Osaka$^{2}$}
\author{Gota Yamaguchi$^{2}$}
\author{Yasuhisa Sano$^{2,9}$}
\author{Kazuto Yamauchi$^{2,9}$}
\author{Jumpei Yamada$^{2,9}$}
\author{Uwe Bergmann$^{7}$}
\author{Matthias F. Kling$^{4,5}$}
\author{Claudio Pellegrini$^{4}$}
\author{Makina Yabashi$^{2,10}$}
\author{Nina Rohringer$^{8,11}$}
\author{Takahiro Sato$^{4}$}
\email{takahiro@slac.stanford.edu}
\author{Agostino Marinelli$^{4,6}$}
\email{marinelli@slac.stanford.edu}

\affiliation{$^1$Department of Advanced Materials Science, The University of Tokyo, Chiba 277-8561, Japan\\
$^2$RIKEN SPring-8 Center, Hyogo 679-5148, Japan\\
$^3$University of Hamburg, Institute for Experimental Physics/CFEL,  Hamburg 22761, Germany\\
$^4$SLAC National Accelerator Laboratory,  California 94025, USA\\
$^5$Department of Applied Physics, Stanford University, Stanford, California 94305, USA\\
$^6$Stanford PULSE Institute, SLAC National Accelerator Laboratory,  California 94025, USA\\
$^7$Department of Physics, University of Wisconsin Madison, Madison, Wisconsin 53706, USA\\
$^8$Center for Free-Electron Laser Science CFEL, Deutsches Elektronen-Synchrotron DESY, Hamburg 22603, Germany\\
$^9$Graduate School of Engineering, Osaka University, Osaka 565-0871, Japan\\
$^{10}$Japan Synchrotron Radiation Research Institute, Hyogo 679-5198, Japan\\
$^{11}${Department of Physics, University of Hamburg, Hamburg 20355, Germany}
}

\begin{abstract}
We present the first direct experimental confirmation of attosecond pulse generation in the hard X-ray regime with a free-electron laser.
Our experiment is based on measurements of a nonlinear optical phenomenon known as
amplified spontaneous emission (ASE) from 3\textit{d} transition metals. 
By analyzing the yield of the collective X-ray fluorescence induced by ultrashort pulses at the Linac Coherent Light Source, we identify the generation of attosecond pulses and shot-to-shot fluctuations in their duration, ranging from 100 as to 400 as.
The observed product of bandwidth and pulse duration for 100 as pulses is approximately 2 fs$\cdot$eV, indicating the generation of nearly transform-limited pulses.
 Our results extend the photon energy reach of attosecond techniques by one order of magnitude, providing the ability to simultaneously probe matter on the time-scales of electronic phenomena and with atomic spatial resolution.
Furthermore, attosecond hard X-ray pulses can outrun the fastest radiation damage processes, paving the way to single-shot damage-free X-ray measurements.
\end{abstract}
\maketitle

\section{Main}\label{sec1}
Attosecond science with ultrafast optical pulses has revolutionized our understanding of electron dynamics by enabling observation of charge migration, electron correlation effects, and field-driven processes on their natural timescales \cite{Krausz2009}. 
In the early days of attosecond science, high harmonic generation (HHG) \cite{Corkum1994} has been the primary tool for generating attosecond pulses \cite{Hentschel2001, Paul2001}. While this technique can produce attosecond extreme ultraviolet pulses, further shortening the wavelength presents significant challenges due to a  reduction in conversion efficiency at higher photon energies.

Free-electron lasers (FELs) are the brightest sources of short-wavelength radiation, with a peak power that exceeds table-top HHG sources and synchrotron radiation sources by several orders of magnitude \cite{Duris2020, Zhirong2007, Pellegrini2016}. 
In FELs, a temporally compressed electron bunch is injected into a periodic array of dipole magnets, called an undulator.
As the electrons traverse the undulator, they undergo oscillatory motion due to the Lorentz force and emit radiation.
The ponderomotive interaction between the electrons and the radiation induces an electron density modulation on the same length-scale as the undulator radiation, leading to coherent amplification of the radiation pulse. FELs typically operate in the self-amplified spontaneous emission (SASE) regime, where the amplification process of radiation is initiated by shot noise in the eletron beam current, and the resulting temporal profile of the pulse is composed by several coherent spikes that are uncorrelated with respect to each other \cite{Pellegrini2016,bonifacio1994spectrum}. If the electron bunch length is comparable to the SASE cooperation length, which is on the order of attoseconds, the corresponding FEL pulses can be composed of a single coherent spike \cite{bonifacio1994spectrum}.

FELs have recently been employed for the generation of attosecond soft X-ray pulses \cite{Duris2020}. This was accomplished with the development of electron bunch shaping techniques that enable the production of a single high-current spike, leading to the emission of isolated attosecond pulses \cite{Duris2020, Zhang2020,Eduard2023,Li2024}. 
Attosecond soft X-ray FELs (XFELs) have demonstrated a remarkable degree of flexibility, with the generation of Terawatt-scale pulses \cite{Franz2024}, two-color pulse pairs for attosecond pump/probe applications \cite{Guo2024}, and controllable pulse trains \cite{duris2021controllable,robles2025spectrotemporal}.
Using these techniques, FELs have extended attosecond nonlinear spectroscopy and the detection of attosecond phenomena to the soft X-ray region \cite{ONeal2020, Driver2024, Guo2024, Alexander2024, li2024attosecond}.  
Similar bunch shaping techniques have also been employed in the hard X-ray region, resulting in the generation of  pulses with broad spectral bandwidth \cite{Marinelli2017, HuangPRL2017, malyzhenkov2020single,Yan2024}.

An obvious challenge in production and utilization of attosecond hard XFEL pulses is their characterization in the time domain.
In the soft X-ray region, the temporal profile of individual attosecond pulses has been measured using angular streaking techniques, where photoelectrons are produced by irradiating a noble gas with soft X-ray pulses and are streaked by the rotating electric-field vector of an infrared circularly polarized pulse
\cite{Hartmann2018, Li2018}. 
In the hard X-ray region, no direct experimental evidence of the attosecond pulse duration has been confirmed yet, and the pulse duration has been inferred by comparing measured spectra to theoretical models and simulations of the FEL process \cite{HuangPRL2017, Marinelli2017, Yan2024}.
However, such estimation has large uncertainties because the spectral width strongly depends on the energy chirp of the lasing part of the electron bunch \cite{Inubushi2012, Krejcik:2017lta}, and the measured spectra only provides a lower but not an upper boundary for the pulse duration.
%For instance, in the case of linearly chirped Gaussian pulses,
%the duration can vary from that of a transform-limited pulse to up to $\sqrt{2}$ of that value \cite{HuangPRL2017}.
%Since actual hard XFEL pulses may exhibit higher-order chirp and shot-by-shot fluctuations in the energy chirp of the electron beam, it remains unclear (1) whether the duration of hard X-ray pulses with a single spectral spike actually reaches the attosecond regime and (2) how much the duration fluctuates from pulse to pulse.

Here, we present the first direct experimental evidence for attosecond hard X-ray pulses.
We propose and demonstrate a new temporal diagnostics technique for attosecond hard X-ray pulses.
Our technique utilizes amplified spontaneous emission (ASE), a form of collective spontaneous emission that has been proposed and intensively studied in the microwave and visible spectral regions \cite{PhysRev.93.99, 1982Gross}.
In the hard X-ray region, this effect can be observed when tightly focused X-ray pulses irradiate 3\textit{d} transition metal targets.
When the pump photon energy exceeds the $K$-shell binding energy and the pump intensity surpasses 10$^{18-19}$ W/cm$^2$, there are short-lived core-excited states that lead to population inversion in the irradiated volume. 
X-ray fluorescence photons spontaneously emitted at the entrance of the target trigger stimulated emission and amplification of the collective spontaneous emission along the pump pulse propagation direction \cite{Rohringer2009, Rohringer2012, Yoneda2015}.
Since the photoabsorption cross section does not strongly depend on the photon energy, the ASE signal is predominantly determined by the temporal pulse shape, pulse energy, and transverse beam profile on the target.
Given that the pulse energy and transverse beam profile on the target can be experimentally evaluated, the pulse duration can be estimated from the ASE signal.
A unique feature of ASE, distinguishing it from other hard X-ray nonlinear effects, is the generation of massive nonlinear signals with the measured number of photons up to $10^8$ \cite{Kroll2018, Doyle:23, Linker2025}.
As long as the amplification remains below saturation, one can evaluate the pulse duration on a shot-by-shot basis by comparing the measured nonlinear signal with the numerical simulation \cite{PhysRevA.109.033725, PhysRevA.109.063705}.
Based on this approach, we characterized the pulse duration of ultrafast hard X-ray pulses produced with the photocathode laser shaping technique (See Methods) at the Linac Coherent Light Source (LCLS) \cite{Emma2010}.

\section{Results}\label{sec2}

\begin{figure}
    \centering
    \includegraphics[width=\linewidth]{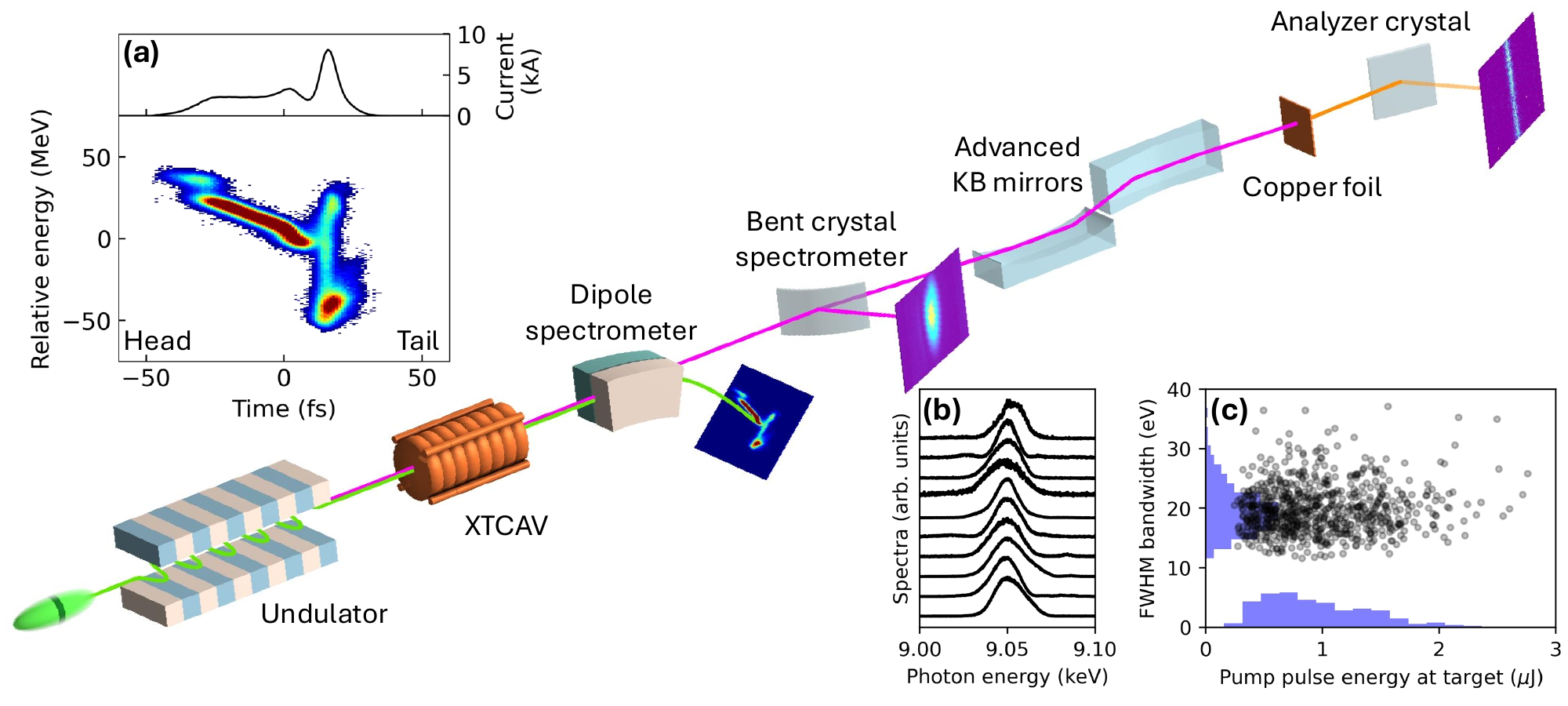}
    \caption{Schematics (not to scale) of the experiment. The pump attosecond X-ray pulses are emitted from a current spike in the electron bunch. The electron beam longitudinal phase space was measured using an X-band transverse deflecting cavity (XTCAV) and dipole spectrometer (a). The pulse is then propagated to the XPP instrument where the spectrum (b) and pulse energy (c)  are characterized in a non-invasive way. The pump pulse is focused with an Advanced KB optics to a 150 nm spot. The nanofocused X-ray pulse irradiates a 20-$\mu$m copper foil, and the produced amplified spontaneous emission is measured with a spectrometer consisting of flat Si (220) analyzer crystal and two-dimensional detector.}
    \label{fig:layout}
\end{figure}

The measurement of ASE was performed at the X-ray Pump-Probe (XPP) instrument \cite{Chollet2015}, as schematically shown in Fig. 1.
The 9.05 keV X-ray pulses were generated from highly compressed electron beam with peak current of $\sim$10 kA (Fig. 1(a)).
The X-ray pulses were focused to a full-width at half maximum (FWHM) of 150 nm spot size using a total-reflection focusing optics system, called the advanced Kirkpatrick-Baez (AKB) mirror system \cite{Inoue2025}.
We positioned a transmissive intensity monitor just upstream of the focusing optics. 
This monitor detected X-ray scattering from a 10-$\mu$m-thick Kapton film using a photodiode.
The shot-by-shot output of the intensity monitor was then correlated with the pulse energy at the focal point, which was measured using a laser power meter calibrated in advance with an X-ray calorimeter \cite{Kato2012}.
The intensity monitor allowed non-invasive and precise evaluation of the pulse energy for each individual XFEL pulse.
The spectrum of the XFEL pulse was measured using a transmissive spectrometer \cite{Zhu2012} based on a thin, bent silicon (220) crystal (10 $\mu$m thickness) and a two-dimensional detector (Alvium), positioned between the intensity monitor and the focusing optics. 
 We found that approximately 6.5\% of all pulses exhibited a single spike in their spectrum (Fig. 1 (b) shows spectrum for the single-spike pulses), which could be well-fitted by Gaussian functions with a coefficient of determination exceeding 99.5\% (see Methods). 
 Figure 1(c) shows the pulse energy at the focal position and the FWHM bandwidth of single-spike pulses. 
 The average pulse energy was 1 $\mu$J, with peaks reaching up to 3 $\mu$J, while the FWHM bandwidth was around 20 eV.
We used a 20-$\mu$m-thick copper (Cu) foil as a target. The target was placed at the focal position and continuously translated to ensure that each pulse irradiated an undamaged surface. X-ray emission spectra in the forward direction were measured with a dispersive spectrometer set in the horizontal plane ($\pi$ polarization) consisting of a flat silicon (220) crystal and a two-dimensional detector with single-photon sensitivity (Jungfrau 1M \cite{Mozzanica2018}).

Figure 2(a) shows the average spectrometer image for single-spike pulses with a pulse energy exceeding 1.0 $\mu$J.
We clearly observed two fluorescence lines at  8048 eV and 8028 eV, corresponding to $K\alpha$1 (2$p_{3/2}$ $\rightarrow$ 1$s$)  and  $K\alpha$2 (2$p_{1/2}$ $\rightarrow$ 1$s$) emissions, respectively.
The intensity of the $K\alpha$1 emission is significantly higher than that of $K\alpha$2 (Fig. 2(b)), which is in stark contrast to the case of spontaneous emission, where the intensity ratio is 2:1.
This result clearly indicates that the observed emission is of nonlinear origin and that an extremely high X-ray intensity on the order of 10$^{19}$ W/cm$^2$ was achieved at the focal position.

We can first roughly estimate the pulse duration from the threshold pulse energy for ASE.
Figure 2(c) shows the average number of detected K$\alpha$1 photons as a function of pulse energy at the target. 
For comparison, data for multi-spike pulses (where the coefficient of determination is between 0.8 and 0.95) and example spectra are also included.
It is evident that significant ASE was not observed for multi-spike pulses, indicating that their peak intensity is lower than that of single-spike pulses for the same pulse energy.
The threshold pulse energy for ASE was around 1.5 $\mu$J for single-spike pulses.
In ASE generation from a 20-$\mu$m-thick Cu target, 
the threshold intensity was experimentally determined to be $2 \times 10^{19}$ W/cm$^2$ for 7-femtosecond X-ray pulses with central photon energy of 9.00 keV \cite{Yoneda2015}.
Assuming that the same intensity threshold applies in the present case, the FWHM pulse duration can be estimated as 
1.5 $\mu$J / (2 $\times 10^{19}$ W/cm$^2$) / (150 nm $\times$ 150 nm) $\sim$ 300 as, although this estimation does not take into account the similarity between the core-hole lifetime of Cu and the pulse duration, as discussed later, nor the shot-by-shot fluctuations in pulse duration.

\section{Discussion}

\begin{figure}
    \centering
    \includegraphics[width=\linewidth]{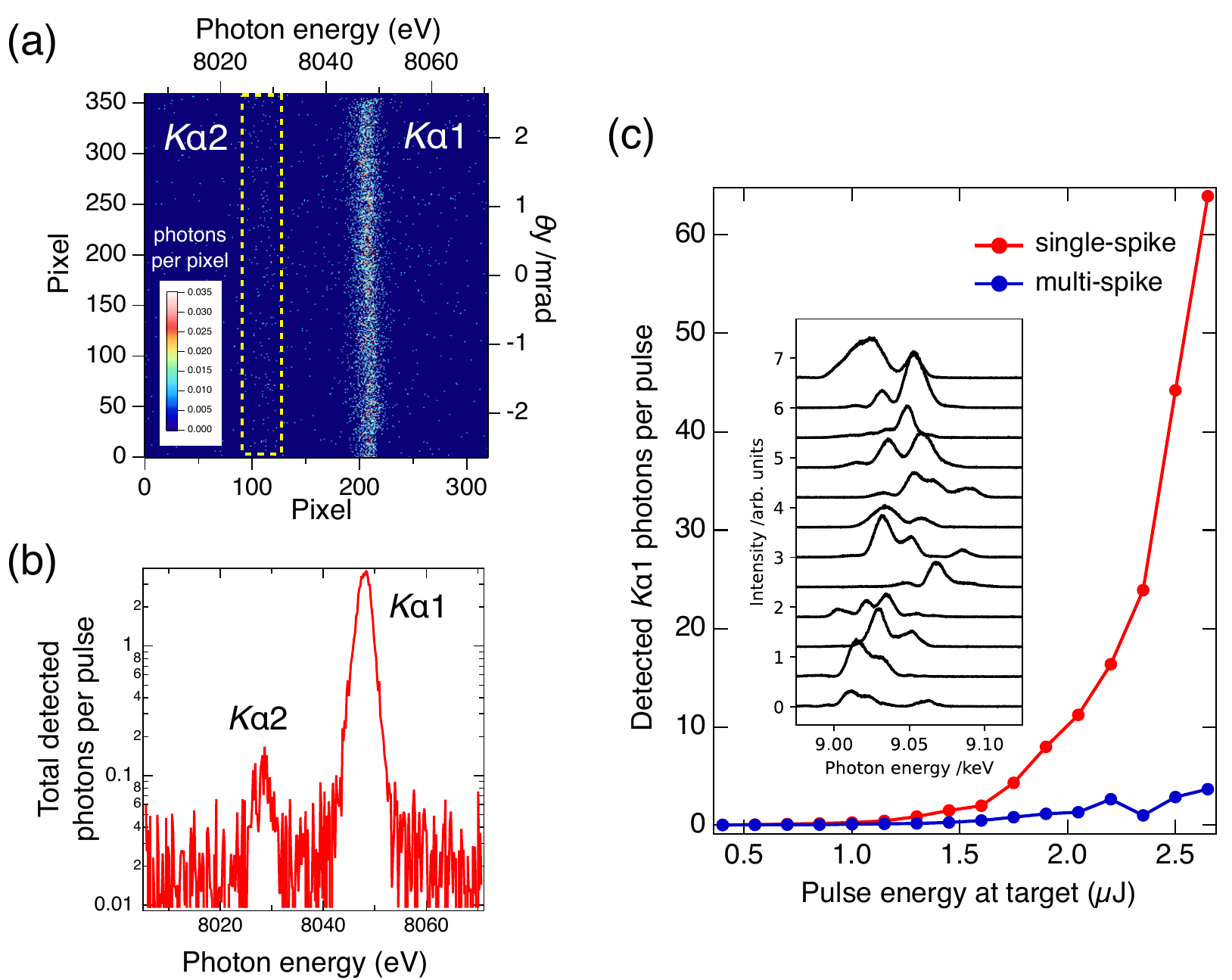}
    \caption{Amplified spontaneous emission from 20-$\mu$m-thick copper foil pumped with nanofocused XFEL pulses. (a) Average detector image of the spectrometer for single-spike pulses with a pulse energy exceeding 1.0 $\mu$J.
    (b) Spectrum of amplified spontaneous emission calculated from (a).
    (c) The average number of detected $K\alpha$1 photons for single-spike pulses (coefficient of determination exceeds 0.99) and multi-spike pulses (coefficient of determination is between 0.8 and 0.95)
    as a function of pulse energy at the target. The inset shows typical single-shot spectra of multi-spike pulses.}
    \label{fig2}
\end{figure}

For a shot-by-shot evaluation of the pulse duration, we compared the measured ASE signal with the simulation results for different pulse durations and pulse energies. The simulation employs comprehensive three-dimensional numerical model that describes the ASE process via a generalized Maxwell-Bloch formalism \cite{PhysRevA.109.033725, PhysRevA.109.063705, Halavanau1} (see Methods). 
In the simulations, we assume that the temporal shape of the pulse has no prominent satellite peaks and can be described by a Gaussian profile.
This assumption of a single peak in the temporal profile is supported by the measured spectra and numerical beam dynamics simulations (see Methods). %Fig. \ref{fig:genesis_sims} in Methods).
%@Alex, I made the sentence a bit milder. I think we do not need to claim the consistensy of the spectral shape for experiment and simulation. 

\begin{figure}
    \centering
    \includegraphics[width=0.75\linewidth]{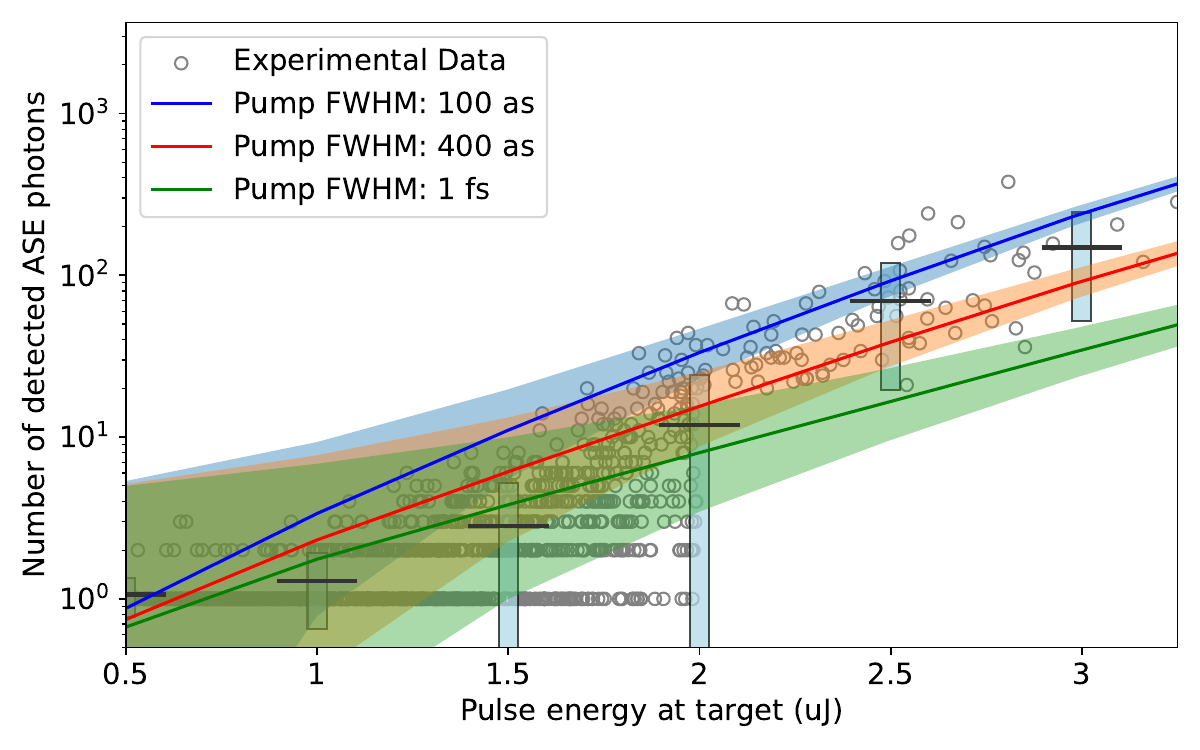}
    \caption{Comparison between measured and simulated number of detected $K\alpha_1$ ASE photons by the spectrometer detector. 
Solid curves represent the expected number of detected photons for different FWHM pulse durations of the incident X-ray pulse (blue: 100 as, red: 400 as, green: 1 fs).
The shaded regions around the curves indicate the 95\% confidence intervals based on Poisson statistics.
Gray horizontal lines and light blue bars represent the average and standard deviation of the detected $K\alpha_1$ photons, grouped by pulse energy ranges of
0.5$\pm$0.25 $\mu$J, 1.0$\pm$0.25 $\mu$J, 1.5 $\pm$0.25 $\mu$J, 2.0$\pm$0.25 $\mu$J, 2.5$\pm$0.25 $\mu$J, 3$\pm$0.25 $\mu$J, respectively.
}
    \label{fig:ASE-sim}
\end{figure}

Figure \ref{fig:ASE-sim} presents the measured ASE signal for single-spike pulses (with a coefficient of determination exceeding 0.99) alongside the simulated ASE signal for pulses with FWHM durations of 100 as, 400 as, and 1 fs, assuming an isotropic Gaussian transverse profile with a FWHM diameter of 150 nm.
Since the expected ASE signal after the analyzer silicon crystal is on the order of a few photons for pulse energies below 2 $\mu$J, significant pulse-by-pulse fluctuations are anticipated due to photon shot noise.
The shaded areas for each simulated curve represent the 95\% confidence interval, assuming a Poisson distribution of the emitted ASE photons.

For pulse energy higher than 2 $\mu$J, the confidence intervals for the three simulation curves are well separated, allowing us to discuss the pulse duration for each pulse.
The measured ASE signal for most pulses is between the simulation results for 100 as and 400 as, clearly indicating (1) the production of attosecond hard X-ray pulses and (2) the presence of pulse-by-pulse fluctuations in the pulse duration.
Since the FWHM bandwidth for these pulses is 15-35 eV (Fig. 1(c)) and the product of the bandwidth and pulse duration for 100 as is close to the value for transform-limited Gaussian pulses, our results indicate the generation of  nearly transform-limited pulses.
It is noteworthy that our simulation of the electron beam dynamics also supports the possibility of generating pulses with a duration of 100 as (see Methods).

For pulse energy less than 2.0 $\mu$J, the confidence intervals of three simulations are overlapped with each other, indicating challenges in shot-by-shot characterization of the pulse duration.
For this pulse energy range, the average number of measured ASE photons  (black bar) roughly agrees with the simulation results for a 1 fs pulse.
Given the potential pulse-by-pulse fluctuations in pulse duration, this suggests that a fraction of the pulses have sub-fs durations.

Finally, we discuss future perspectives on pulse duration measurement using ASE.
In the current experiment, the accuracy of the estimated pulse duration is primarily limited by the restricted signal level.
Since most of the ASE photons do not satisfy the Bragg condition of the silicon analyzer crystal, they do not reach the spectrometer detector (see Methods). 
Measuring the net ASE signal without a spectrometer would significantly improve the accuracy of pulse duration characterization.
Although ASE is emitted in the forward direction and spatially overlaps with the transmitted pulse, using an appropriate two-dimensional detector with sufficient photon energy resolution would enable the measurement of the net ASE signal.
It may be possible to correlate the precise pulse duration with accelerator and electron beam parameters. Since many of these parameters can be monitored without interfering with XFEL pulse generation, data mining could enable non-destructive estimation of the pulse duration. This approach would be highly beneficial for the practical utilization of attosecond hard X-ray pulses.

Another intriguing subject is extending the applicability of this diagnostic technique to shorter pulse durations. The ASE signal becomes insensitive to pulse duration when the pulse duration is much shorter than the core-hole lifetime of the target atoms.
For example, when using copper (with a core-hole lifetime of 400 as \cite{Jurek:zd5003, CAMPBELL20011}) as a target, the ASE signal does not depend on the pulse duration when it is shorter than 50 as (see Methods).
The use of ASE from $L$-shell holes in atoms with higher atomic numbers will enable temporal diagnostics of hard X-ray pulses with durations of a few tens of attoseconds or less.

%Ichiro: I feel this sentence is too detailed. If we claim this, we should add propoer experimental data and simulation results, which will disperse the main story.
%Although the actual transverse shape of the XFEL attosecond pulses is expected to also be close to a Gaussian, the nonlinearities in the electron beam compression and dynamics may lead to some degree of transverse inhomogeneity. Nevertheless, we argue that when averaged over an ensemble of shots, the transverse mode shape fluctuations do not affect the obtained results. 

In summary, we propose and demonstrate a temporal diagnostic technique for hard X-ray attosecond pulses using ASE. The application of this technique to hard X-ray pulses produced with
the photocathode laser-shaping technique at the LCLS revealed the generation of attosecond X-ray pulses and pulse-by-pulse fluctuations in pulse duration. 
Our measurement suggests that some of the pulses have a duration of 100 as, which corresponds to nearly transform-limited pulses.
The attosecond hard X-ray pulses confirmed by our technique pave the way for a new paradigm in X-ray measurements, such as electronic damage-free measurements. Although existing femtosecond hard XFELs enable measurements before significant X-ray-induced atomic displacements \cite{Chapman2011, Inoue2016, Inoue2022}, a key remaining challenge is mitigating electronic damage. This damage is primarily caused by secondary ionization triggered by photoelectrons, which occurs on a timescale of a few femtoseconds \cite{Ziaja2005, Inoue2021, Inoue2023}. The attosecond hard X-ray pulses confirmed in this study can significantly reduce electronic damage, unlocking new scientific opportunities, such as nonlinear spectroscopy with intense X-ray pulses and damage-free visualization of valence electrons through diffraction techniques.

\section{Methods}\label{sec11}

%\subsection{Generation of hard X-ray attosecond XFEL pulses}
\subsection{Generation of attosecond hard XFEL pulses}

\begin{figure}[h!]
    \centering
    \includegraphics[width=0.5\linewidth]{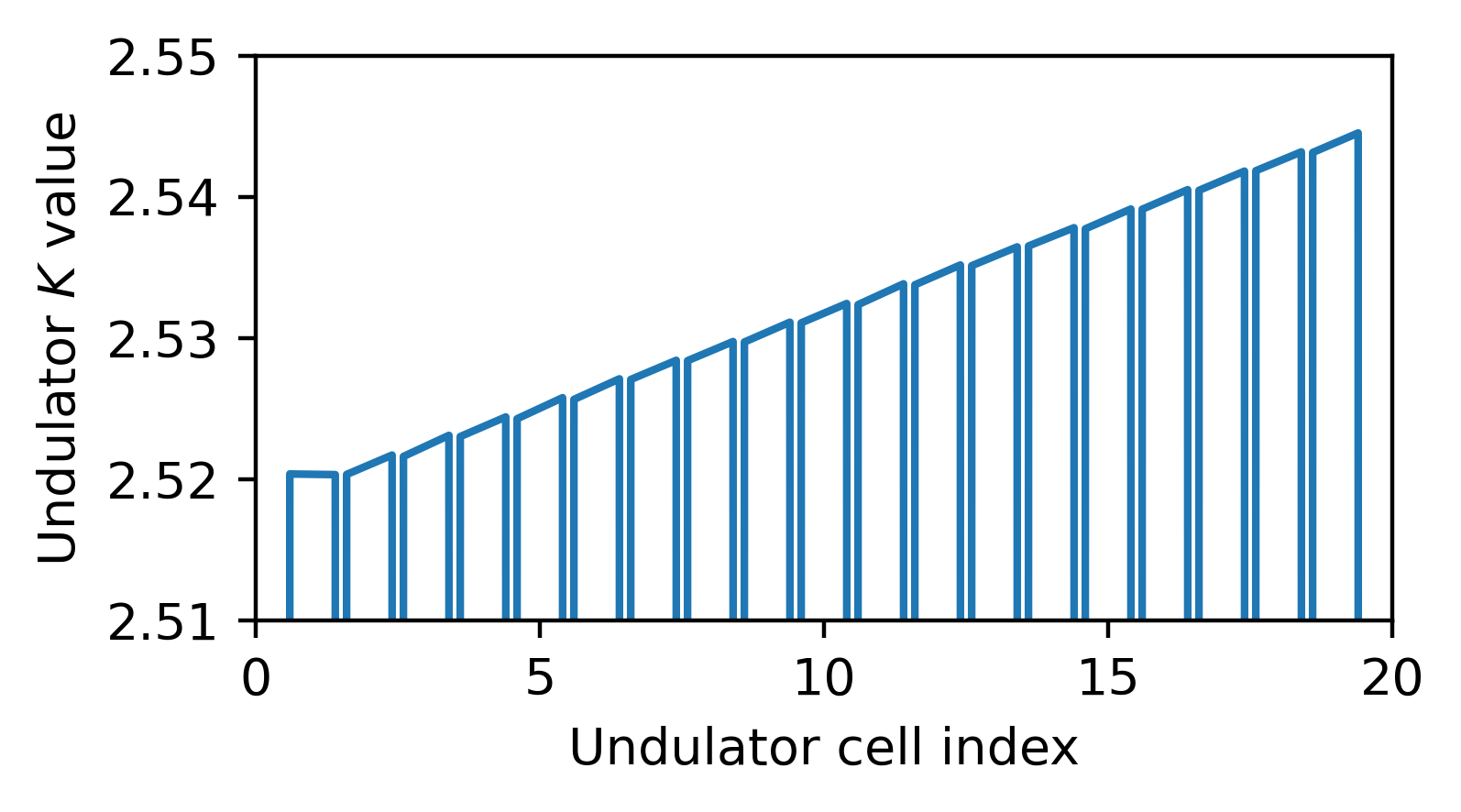}
    \caption{Peak undulator $K$ values for the lasing undulators. The rest of the LCLS undulators were scrambled to prevent lasing.}
    \label{fig:exp_und_Ks}
\end{figure}

The hard X-ray attosecond pulses used in these experiments were generated using the photocathode laser shaping technique \cite{Zhang2020}. The electron beam dynamics are similar to earlier experiments at the LCLS in the soft X-ray regime. The LCLS photoinjector laser pulse is composed of two roughly gaussian temporal parts. In standard (SASE) operation, they are stacked with a time delay such that the laser profile hitting the cathode has a flat-top shape. In this experiment, we separated the two laser pulses to generate a small dip in the temporal profile. This small dip in the electron bunch current seeds the microbunching instability, whereby the small variation in current leads to an energy modulation due to longitudinal space charge. It is, in turn, converted back into a stronger density modulation by the two electron bunch compressors. The energy modulation is compressed one final time by adjusting the momentum compaction of the magnetic dogleg (electron beam transport line where the beam is deflected twice in opposite directions), one hundred meters upstream of the LCLS undulators. This leaves a sub-fs spike in the beam with estimated 10-20 kA peak current. The short, high current spike supports lasing with a pulse duration below 1 fs FWHM. In addition, longitudinal space charge forces between the dogleg and the end of the undulators impart a strong linear energy chirp within the spike \cite{Zhang2020}, with estimated magnitude on the order of 100 MeV/fs on top of the 10.263 GeV beam energy. These values are corroborated by the x-band transverse deflecting cavity (XTCAV) measurements \cite{Krejcik:2017lta} shown in Figure~\ref{fig:layout}. We note that the XTCAV temporal resolution is insufficient to resolve the structure inside of the spike. The energy chirp can be compensated with a linear taper of the undulator strength. Generally, for a linear energy chirp $\dot{\gamma}$ the appropriate rate of change of the undulator strength $K$ is:
\begin{equation}
    \frac{dK}{dz}=\frac{2+K_0^2}{\gamma K_0}\frac{\lambda_r}{\lambda_u}\frac{\dot{\gamma}}{c},
\end{equation}
 where $K_0$ is the nominal undulator strength, $\gamma$ is the beam Lorentz factor, $\lambda_r$ is the radiation wavelength, and $\lambda_u$ is the undulator period. In this experiment, we used a rate of $\Delta K=0.0014$ per undulator segment, corresponding to a roughly 75 MeV/fs chirp. The undulator taper profile is shown in Figure~\ref{fig:exp_und_Ks}. The linear undulator taper simultaneously compensates the energy chirp in the spike, allowing it to lase and generate a shorter pulse, and suppresses lasing outside of the spike where the energy chirp is much smaller.

\subsection{Coefficient of determination}
We used the coefficient of determination, $R^2$, as a measure of how well a single Gaussian function describes each single-shot spectrum.
In our experiment, we measured single-shot spectra in the photon energy range of 8.95 keV to 9.15 keV with a sampling step of 0.1 eV (total number of photon energy points was $n=2001$).

Let $y_i$ and $\hat{y_i}$ represent the observed and predicted intensities for $i$th photon energy, respectively.
The coefficient of determination is then defined as
\begin{equation}
R^2=1-\frac{\sum_{i=0}^{n} (y_i-\hat{y_i})^2}{\sum_{i=0}^{n} (y_i-\bar{y})^2},
\end{equation}
where $\bar{y}$ is the mean observed intensity over all measured photon energy points.

When the observed spectrum is well described by a single Gaussian pulse, 
$R^2$ approaches unity.
Conversely, when the spectrum contains multiple peaks and deviates from a single Gaussian shape, $R^2$ decreases, reflecting a poorer fit to the model.

\subsection{Numerical simulations of the electron beam dynamics}
\begin{figure}[h!]
    \centering
    \includegraphics[width=0.6\linewidth]{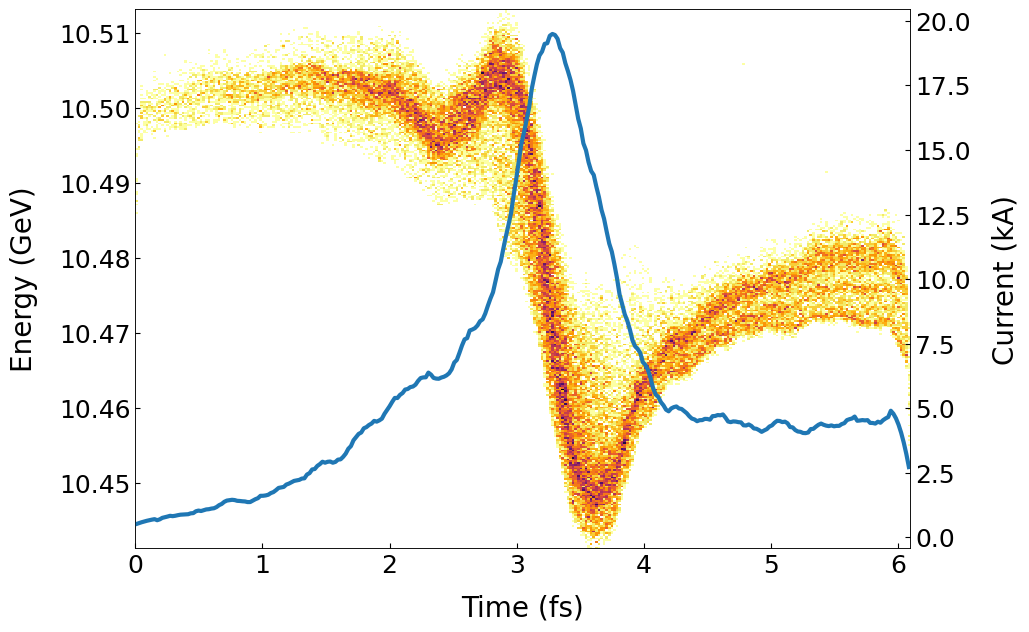}
    \caption{Simulated electron beam longitudinal phase space at the entrance of the undulator (with Elegant tracking code). The electron beam has a peak-to-peak energy spread of 72 MeV and a peak current of 20 kA.}
    \label{fig:beam_sims}
\end{figure}

\begin{figure}[h!]
    \centering
    \includegraphics[width=\linewidth]{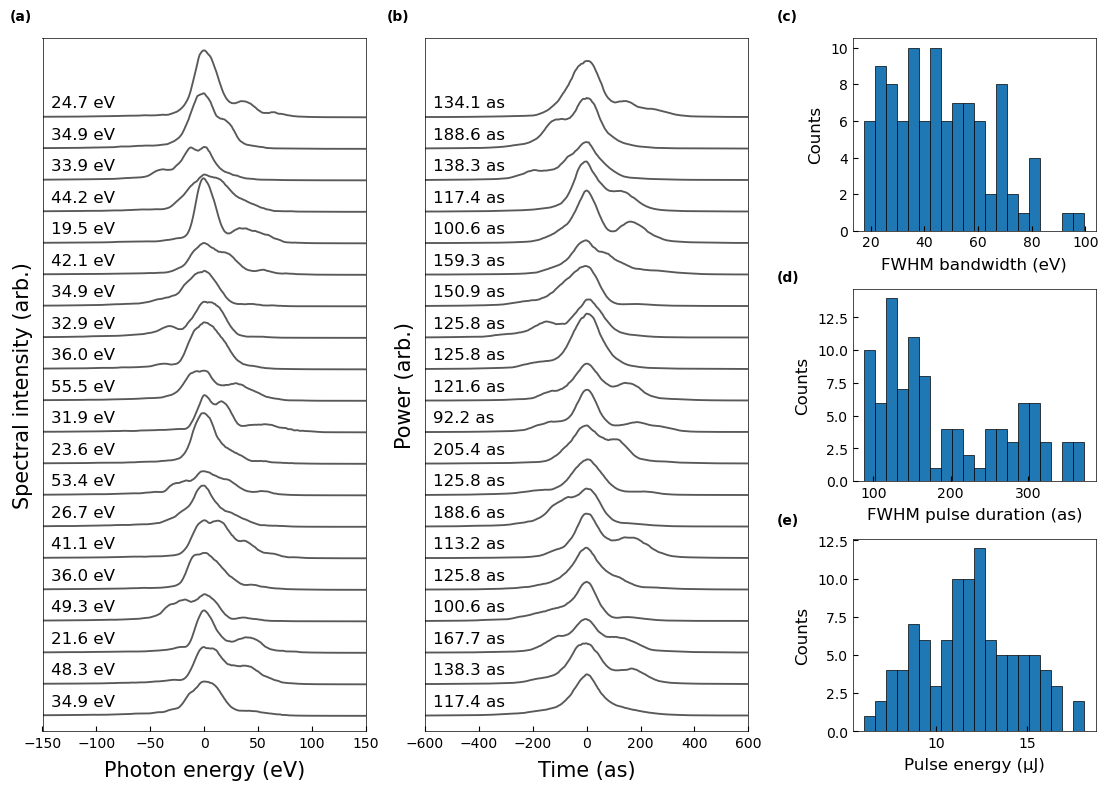}
    \caption{Simulated spectral (a) and temporal (b) LCLS hard X-ray attosecond pulses, and the corresponding pulse statistics of FWHM bandwidth (c), pulse duration (d) and pulse energy (e).}
    \label{fig:genesis_sims}
\end{figure}

In order to numerically validate the assumption of a single peak in the temporal profile of the pulse and the feasibility of generating $100$ as pulses at the LCLS, we have performed start-to-end simulations. We begin with the electron beam dynamics tracking from the photoinjector with a cathode laser stacker delay leading to a dip in the beam current profile. We use the Astra particle-in-cell (PIC) code for injector simulations \cite{flottmann_astra} and Elegant for tracking the particles through the rest of the accelerator \cite{borland2000elegant,wang2006pelegant}. As a result, we obtain an ultra-relativistic beam with a current spike shown in Figure~\ref{fig:beam_sims}, with $20$ kA scale peak current and sub-femtosecond duration. The peak-to-peak energy spread within the spike is around $75$ MeV, consistent with the experimental XTCAV measurements. Additionally, at the entrance of the undulator, the beam has a roughly linear energy chirp within the spike of magnitude $100$ MeV/fs. That linear chirp is compensated by a linear undulator taper of similar magnitude to the one used in the experiment. We then simulate FEL lasing in Genesis 1.3 v4 \cite{reiche1999genesis,genesis4github} code. As in the experiment, we reverse-taper the undulators linearly to compensate the linear energy chirp in the electron bunch, and perform 100 statistically independent simulations starting from random shot noise. The results of the FEL simulations are summarized in Figure~\ref{fig:genesis_sims}. Figure~\ref{fig:genesis_sims} (a) and (b) show a selection of single shot spectra and power profiles, respectively. The simulated temporal profiles exhibit a prominent single peak without significant satellite structures (Fig.~\ref{fig:genesis_sims} (b)), supporting our assumption that the pulse shape can be approximated by a Gaussian function.
Figure~\ref{fig:genesis_sims} (c), (d), and (e) show histograms of the FWHM bandwidth, pulse duration, and pulse energy observed in the simulation. We note that outliers with very large bandwidths are cases where the pulses have multiple spikes and therefore the FWHM is a less useful metric. The pulse energy shown in Fig.~\ref{fig:genesis_sims} (e) is that obtained just at the end of the undulator. These simulations confirm the possibility of generating pulses with $20-30$ eV spectral bandwidth and $100$ attosecond duration with $\mu$J-scale pulse energy at LCLS, which is consistent with our pulse duration characterization based on ASE measurement. Due to intrinsic LCLS pulse energy jitter, especially in the attosecond mode, we sampled various pump pulse energies in one dataset.

\subsection{Numerical simulations of amplified spontaneous emission}
\begin{figure}[h!]
    \centering
    \includegraphics[width=0.75\linewidth]{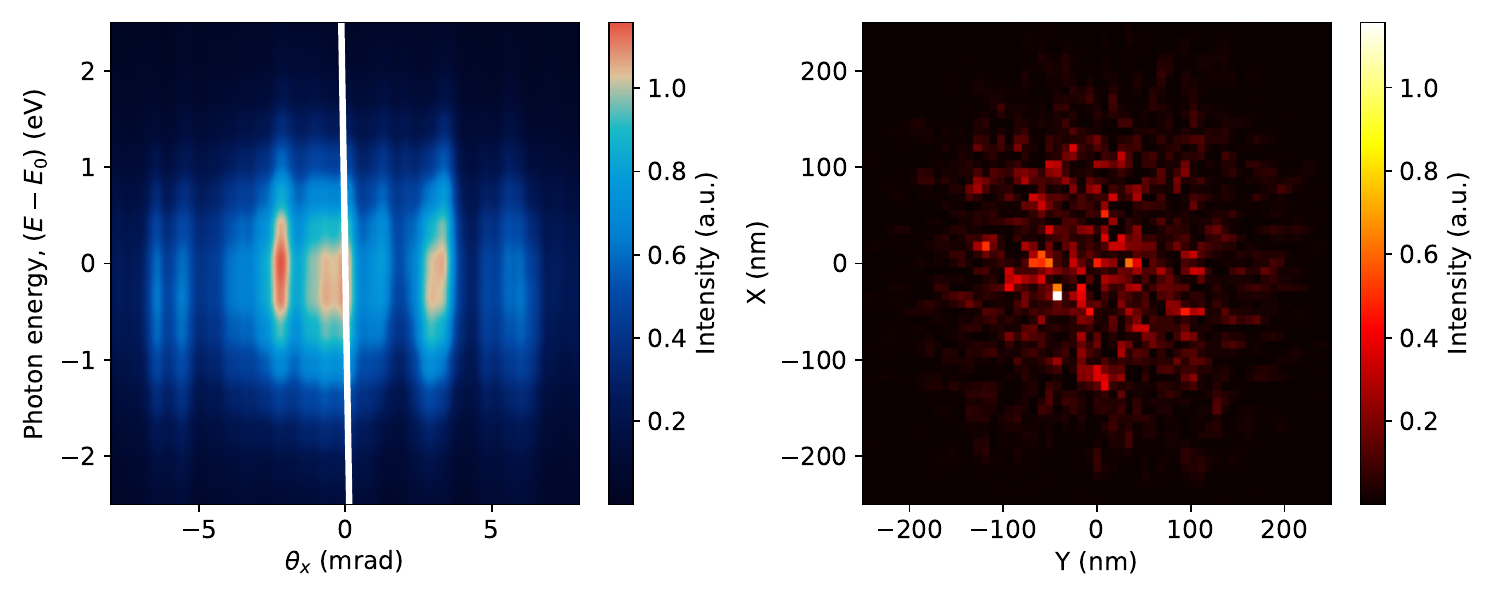}
    \caption{Spectral-angular  and transverse intensity distribution of the simulated ASE intensity at the exit of the gain medium for the XFEL pump duration of 300 as. The white line corresponds to the spectral-angular acceptance window along which the photons were reflected into the 2D detector.}
    \label{fig:DuMond}
\end{figure}

Simulation of ASE generation is performed in state-of-the-art three-dimensional Maxwell-Bloch solver, with stochastic terms described in \cite{PhysRevA.109.033725, PhysRevA.109.063705,Linker2025, Halavanau1}.
In our model, we use the paraxial approximation to propagate the emitted $K\alpha1$ radiation. Each numerical realization can be interpreted as a statistically independent single event of collective spontaneous emission.
The copper medium is assumed to be uniform with the concentration of copper atoms of 84 atoms/nm$^3$.
The 3D ASE field amplitudes are propagated through the medium and to the far field using traditional Fourier optics methods. After accounting for the reduction of the field amplitude due to absorption by Kapton windows and air between the target and the detector in the experiment, the spatial intensity profile of the ASE after diffraction by a flat Si (220) analyzer crystal is calculated (Fig. \ref{fig:DuMond}). 
Here, we note that most ASE photons are absorbed by the crystal, as only a limited portion of those emitted within the angular range of $\sim 25$ $\mu$rad for $\sigma$-polarization and $\sim 17$ $\mu$rad for $\pi$-polarization can be diffracted (white line shown in Fig. 7 represents the spectral-angular acceptance window for ASE with $\pi$-polarization), while the overall ASE divergence is on the order of 10 mrad. The flat crystal analyzer accepts up to 5.2 mrad of divergence in the vertical plane, and has the energy acceptance greater than 50 eV, fully covering emitted spectrum.  
%The analyzer losses impact the correlation of the ASE pulse energy with the pump pulse spectra, therefore the comparison with numerical simulations can only be performed in a statistical sense. 
The diffracted field intensities are integrated over the angular range corresponding to that covered in the experiment to estimate the total number of ASE photons detected by the spectrometer detector. We repeated the simulation under different pulse duration conditions and compared the results with the experimental data, as shown in Fig. \ref{fig:ASE-sim}. We performed simulations with different grid resolutions and time steps, optimizing the discretization parameters to ensure that the simulation results are stable and do not change significantly with further refinement.

\subsection{Limitation of pulse duration diagnostics based on amplified spontaneous emission}
The ASE signal becomes insensitive to pulse duration when the pulse duration is much shorter than the core-hole lifetime of the target atoms.
Figure 8 shows simulation results for the expected number of detected $K\alpha_1$ ASE photons from a 20-$\mu$m-thick copper foil in the experimental geometry, for different pump pulse durations. The simulations were performed following the same procedures described above.
For pump pulse durations longer than the core-hole lifetime of copper (400 as \cite{Jurek:zd5003, CAMPBELL20011}), the ASE signal at a fixed pulse energy strongly depends on the pulse duration. Therefore, the duration of the incident pulse can be inferred from the measured ASE signal.
In contrast, the simulated ASE signals for 25 as and 50 as pulses are nearly identical. This indicates that when the pulse duration is much shorter than the core-hole lifetime, it becomes difficult to accurately determine the pulse duration and only an upper limit can be estimated.
Using core-hole states with shorter lifetimes, such as $L$-shell holes in higher-$Z$ elements, could improve the temporal resolution of this method.

\begin{figure}[h!]
    \centering
    \includegraphics[width=0.65\linewidth]{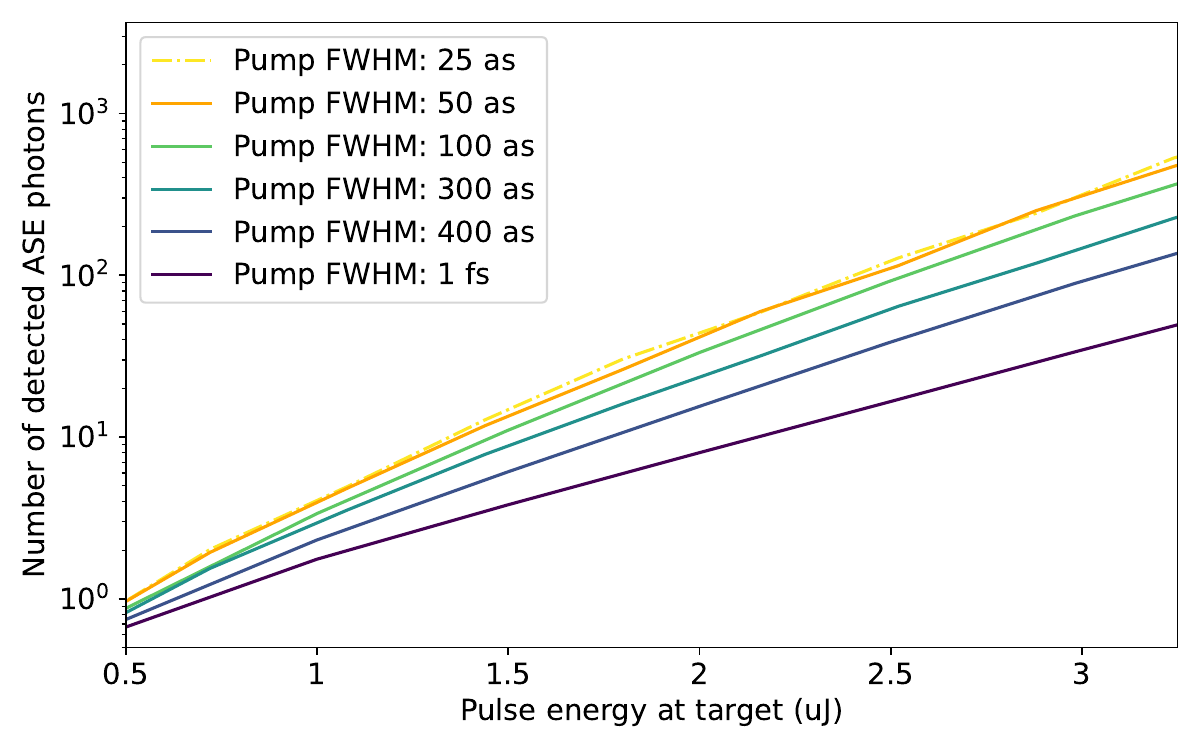}
    \caption{Expected number of detected $K\alpha 1$ ASE photons from 20-$\mu$m-thick copper foil by the spectrometer detector for different durations of the incident pulses. 
    The simulations were performed using the same procedures as those used for Fig. 3.}
    \label{ka1}
\end{figure}

%We also performed various convergence studies to validate the numerical results. %The intrinsic radiation statistics of ASE will be the subject of a future study. We note that in the exponential ASE gain regime, like in our experiment, the transverse profile of ASE consists of individual speckles, which correspond to individual transverse modes (see Fig. \ref{fig:DuMond}). These speckles are eventually eliminated deeper in the exponential gain regime or in saturation \cite{PhysRevA.109.033725, Linker2025}.

\section{Acknowledgements}
We thank  Kenji Tamasaku for careful reading of the manuscript, Jiawei Yan, Ulrike Boesenberg, Stephan Kuschel, Robert Radloff, Rasmus Bunchen, Jan Niklas Leutloff, Taisia Gorkhover, and Yuichi Inubushi, for fruitful discussions and for their interest in this work.

\section*{Data availability}
All data in this study are available from the corresponding author on reasonable request.

\section*{Author contribution}
I.I. conceived the idea of using ASE for pulse duration diagnostics, led the experiment, and conducted data analysis.
R.R., A.H., V.G., and A.M. led the electron beam dynamics simulations, developed the attosecond pulse shaping technique and led machine tuning of attosecond hard X-ray pulses.
D.C., Y.D., V.E., P.F., N.S.S., and Z.Z. also supported the machine tuning.
A.H., A.B., S.C., T.L., and N.R. performed the Maxwell-Bloch simulations to evaluate the pulse duration.
T.S. led the design of the experimental setup for measuring amplified spontaneous emission from copper.
Y.Sano, K.Y., and M.Y. contributed to the development of the optics used in this experiment.
I.I., T.S., D.Z., Y.Sun, M.S., and J.Y. performed the ASE measurements.
U.B. and M.K. contributed to the discussion on data interpretation and scientific applications.
I.I., R.R., A.H., V.G., and A.M. wrote the original draft of the paper, and all authors contributed to improving the manuscript.

\section*{Funding}
This research is supported by JSPS KAKENHI (Grant Numbers: 22KK0233, 23K17149, 23K25131, 24K21199) and JST Precursory Research for Embryonic Science and Technology (JPMJPR24J1). This work at LCLS is supported by the U.S. Department of Energy Contract No. DE-AC02-76SF00515 and by the Basic Energy Science FWP No. 100617 and No. DE-SC0063.
Computer resources for simulations were provided by the National Energy Research Scientific Computing Center (NERSC), a U.S. Department of Energy Office of Science User Facility located at Lawrence Berkeley National Laboratory, operated under Contract No. DEAC02-05CH11231 using NERSC award ERCAP0020725. R.R. acknowledges support from the Stanford University Siemann Fellowship. 

\section*{Competing interests}
The authors declare no competing interests.

\bibstyle{natbib}
\bibliography{sn-bibliography}

\end{document}